\documentclass[reprint,superscriptaddress, amsmath,amssymb, aps,pra,floatfix,]{revtex4-2}

\usepackage[utf8]{inputenc} 
\usepackage[T1]{fontenc}    
\usepackage{hyperref}       
\usepackage{url}            
\usepackage{booktabs}       
\usepackage{amsfonts}       
\usepackage{nicefrac}       
\usepackage{microtype}      
\usepackage{lipsum}		
\usepackage{amsthm} 
\usepackage{graphicx}
\usepackage{natbib}

\usepackage[ruled,vlined,linesnumbered]{algorithm2e}

\newcommand{\ten}[1]{\mathcal{#1}}
\newcommand{\mat}[1]{\mathbf{#1}}
\renewcommand{\vec}[1]{\mathbf{#1}}

\newcommand{\HAT}[1]{\widetilde{}}

\usepackage[utf8]{inputenc}
\usepackage[T1]{fontenc}
\usepackage{graphicx} 
\usepackage{amsmath}
\usepackage{amssymb}
\usepackage{amsthm}
\usepackage{mathtools}
\usepackage[ruled,vlined,linesnumbered]{algorithm2e}
\usepackage[dvipsnames]{xcolor}
\usepackage{tikz}
\usepackage{siunitx}
\usepackage{stmaryrd}
\usepackage{fancyhdr}
\usepackage{titlesec}
\usepackage{secdot}
\usepackage{hyperref}
\usepackage{algpseudocode}
\usepackage{lineno}
\usepackage{todonotes}

\usepackage{booktabs}
\usepackage{geometry}
\newcommand{\qbar}[0]{\vec{\bar{q}}}

\usepackage{lipsum}
\usepackage{amsthm}

\begin{document}

\preprint{APS/123-QED}

\title{Breaking the Curse of Dimensionality: Solving Configurational Integrals for Crystalline Solids by Tensor Networks}
\author{Duc P. Truong}
\affiliation{Los Alamos National Laboratory, Los Alamos, NM 87545, USA}

\author{Benjamin Nebgen}
\affiliation{Los Alamos National Laboratory, Los Alamos, NM 87545, USA}

\author{Derek DeSantis}
\affiliation{Los Alamos National Laboratory, Los Alamos, NM 87545, USA}


\author{Dimiter N. Petsev}
\affiliation{Department of Chemical and Biological Engineering, and Center for Microengineered Materials, 
University of New Mexico, Albuquerque, NM 87131, USA}

\author{Kim {\O}. Rasmussen}
\affiliation{Los Alamos National Laboratory, Los Alamos, NM 87545, USA}

\author{Boian S. Alexandrov}
\affiliation{Los Alamos National Laboratory, Los Alamos, NM 87545, USA}

\date{\today}

\begin{abstract}
Accurately evaluating configurational integrals for dense solids remains a central and and difficult challenge in statistical mechanics of condensed systems. Here, we present a novel tensor network approach that reformulates the high-dimensional configurational integral for identical particle crystals into a sequence of computationally efficient summations. We represent the integrand as a high-dimensional tensor and apply tensor-train (TT) decomposition together with a custom TT-cross interpolation. This approach circumvents the need to explicitly construct the full tensor, which would otherwise be computationally intractable. We introduce tailored rank-1 and rank-2 schemes optimized for sharply peaked Boltzmann probability densities, typical for identical-particle crystals. When applied to the calculation of internal energy and pressure–temperature curves for crystalline $Cu$ and $Ar$ as well as the $\alpha-Sn$ to $\beta-Sn $ phase transition diagram, our method accurately reproduces molecular dynamics simulation results using tight-binding, machine learning (HIP-NN), and MEAM potentials—all within seconds of computation time.
\end{abstract}

\maketitle


%
\textit{Introduction}---For over a century, physicists have sought to rigorously elucidate the emergence of macroscopic properties of matter from the underlying microscopic dynamics of atoms and molecules. This foundational link is formalized by statistical mechanics, which provides a systematic framework for deriving thermodynamic behavior from the statistical treatment of microscopic degrees of freedom. The foundations of statistical mechanics were established through the seminal contributions of Maxwell \cite{Maxwell}, Boltzmann \cite{Boltzmann}, and Gibbs \cite{Gibbs}. It was Gibbs who introduced the canonical partition function in statistical mechanics, a central construct that characterizes a system of \( N \) particles in three-dimensional space using a statistical ensemble formulation:  
\begin{equation}
    Q_N = \frac{1}{N!\Lambda^{3N}}Z_N, \; Z_{N} = \int_{\mathbb{R}^{3N}} e^{- \beta U(\vec{q})} d\vec{q}, 
    \label{eqn:Partition}
\end{equation}
where \( \vec{q} = [x_1, y_1, z_1, \dots, x_N, y_N, z_N] \) are the particle coordinates, and  \( U(\vec{q}) \) is the multi-particle interaction potential. The parameter $\Lambda = h/(2\pi mk_BT)^{1/2}$ is the de Broglie wavelength, where $h$ is Planck's constant, $k_B$ is Boltzmann's constant, $\beta = 1/k_BT$ where T is the temperature, and $m$ is the particle mass. The quantity $Z_N$ is the configurational integral, which accounts for the interactions among particles (atoms and/or molecules) via their potential energy. The evaluation of the configurational integral in systems with strong interparticle interactions remains one of the central challenges in the statistical mechanics of condensed matter.

The physical interpretation of the partition function and the configurational integral in Eq. \eqref{eqn:Partition} is conceptually straightforward: it represents a summation over all configurations compatible with the system’s intermolecular interactions and macroscopic thermodynamic constraints. However, the mathematical complexity escalates dramatically in condensed-phase systems, such as liquids and solids, where interactions between atoms or molecules dominate system behavior. In macroscopic systems, the number of constituent particles can exceed $10^{23}$ making the computation of total potential energy exceedingly challenging. In addition, the configuration integral $Z_N$ (see Eq. \eqref{eqn:Partition}) is $3N$-fold, and traditional numerical techniques are not well-suited to deal with such large-scale problems. For example a simple midpoint numerical integration approach to solve a 100-fold integral (dimensionality $d = 100$) on $n + 1$ nodes in each direction requires $n^d$ function evaluations. This exponential increase in computational complexity is known as the `curse of dimensionality'~\cite{bellman1966dynamic}. Consider the example of $n = d = 100$ integration nodes. This choice requires $10^{200}$
function evaluations and similar number of floating-point operations. Such a computation would need at least $2\times10^{188}$
seconds on a teraflop computer capable of $10^{12}$ floating point operations per second. This time is roughly equal to $4.58\times10^{170}$ times the age of the Universe
\cite{alexandrov2023challenging}. To address these computational challenges, researchers have employed approximate integral equation theories \cite{McQuarrie} or turned to simulations such as Molecular Dynamics (MD) and Monte Carlo (MC) \cite{AllenTildesley}, each constrained by inherent limitations in accuracy, timescale, or convergence.

Recent advances with tensor networks \cite{cichocki2016tensor}, notably the Tensor Train (TT) decomposition, offer a breakthrough in handling high-dimensional integrals \cite{oseledets2010tt,khoromskaia2013tensor, dolgov2020parallel,alexandrov2023challenging}. The TT decomposition transforms previously intractable computational challenges into manageable, practical solutions by dramatically reducing complexity. This shift in numerical techniques, which we exploit here, promises to transform the approach to large-scale problems in physics and chemistry by resolving efficiency issues inherent in conventional methods. 


\textit{Tensor Formulation of $Z_N$}---The integrand in \( Z_N \) is a \( 3N \)-dimensional function \( f(\vec{q}) := e^{-\beta U(\vec{q})} \). To solve the integral, this function must be numerically evaluated in a set of $n^{3N}$ grid points, where $n$ is the number of points in each dimension. This evaluation yields the Boltzmann factor - the array \( \mathcal{F}(i_1, \ldots, i_{3N}) \). The utilization of the cubature rule approximates \( Z_N \) by weighted summations over all dimensions give us:
\begin{equation}
\label{eq:cubature}
Z_N \approx \sum_{i_1 = 1}^{n} w_{i_1} \cdots \sum_{i_{3N} = 1}^{n} w_{i_{3N}} \, \mathcal{F}(i_1, i_2, \dots, i_{3N}).
\end{equation}
It is important to note that the sheer size of the tensor \( \mathcal{F} \) makes it impossible to construct explicitly.
\noindent \textit{Tensor Train (TT) Decomposition:} TT decomposition provides an approximation that compresses high-dimensional tensors, formatting them as a product of small three-dimensional (3D) tensor cores, ordered in a linear chain \cite{oseledets2011tensor}. Below, we adopt a slightly modified TT format of the Boltzmann factor $\ten{F}$, 
\begin{equation}
\label{eq:TT_def}
\begin{split}
\ten{F}(i_1,\ldots,i_{3N}) = \sum_{\vec{\alpha}}^{\vec{r}} 
\ten{G}_1(1,i_1,\alpha_1) \mat{M^1}(\alpha_1, \alpha_1)\\ 
\ten{G}_2(\alpha_1, i_2, \alpha_2) \mat{M^2}(\alpha_2, \alpha_2) \cdots \\
\cdots \mat{M^{3N-1}}(\alpha_{3N-1}, \alpha_{3N-1}) \ten{G}_{3N}(\alpha_{3N-1},i_{3N},1) + \varepsilon. \nonumber
\end{split}
\end{equation}
The components of the vector $\vec{r} = [r_1,\ldots,r_{3N-1}]$, which are the ranges of the indices $\alpha_i$, are called TT ranks, and $\varepsilon$ is the approximation error of the decomposition. Each of the 3D core-tensors, $\ten{G}_k$, is uniquely associated with a single index of the original tensor $\ten{F}$, establishing the TT decomposition as a discrete manifestation of separation of variables.
\textit{Computing $Z_N$ from TT Format}---Once the tensor $\ten{F}$ is in TT-format, we can calculate $Z_N$ by, 
\[
Z_N \approx \prod_{k=1}^{3N}\mat{R_k}, 
\]
where: $\mat{R_1}$ is a row vector of size $r_1$; $\mat{R_k},\ k=2,\ldots,3N-1$ are matrices of size $r_{k-1}\times r_{k}$; 
\begin{equation}
\label{eq:TT - cubature}
\begin{split}
&\mat{R_k} = \sum_{i_k = 1}^{n_k} w^k_{i_k}\ten{G}_k(:,i_k,:)\mat{M^k},
\end{split}
\end{equation}
Hence, the $3N$-dimensional integral $Z_N$ is reduced to $3N$ one-dimensional sums. The challenge is to decompose $\ten{F}$ without accessing all its elements.

\textit{TT-cross Interpolation}---TT-cross is an algorithm for TT decomposition, needed when accessing all entries in a tensor $\ten{F}$ is computationally impossible~\cite{oseledets2010tt}. Specifically, instead of performing consecutive Singular Value Decompositions (SVDs) and reshapings of the unfolding of the tensor $\ten{F}$ (as in the TT-SVD algorithm~\cite{oseledets2011tensor}), TT-cross adaptively samples a small subset (arguably the most important) of $r$ columns and $r$ rows of the unfolding, based on the ideas of pseudoskeleton~\cite{goreinov1997theory} and $\mat{CUR}$ \cite{mahoney2009cur} decompositions. In this way TT-cross
selects a submatrix with maximum volume (i.e., maximum absolute determinant), which ensures an optimal accuracy \cite{goreinov1997theory, oseledets2011tensor} of the approximation. Since computing the exact submatrix with the maximum volume is NP-hard,  practical implementations rely on the heuristic quasi-maxvol algorithms \cite{savostyanov2014quasioptimality}, see \textit{Appendix A} in End Matter.

\textit{TT-cross of $Z_N$}---TT-cross interpolation circumvents the need for full access to the tensor $\ten{F}$'s elements. 
However, the Boltzmann factor poses significant challenges to TT-cross:
(i) Since most of the tensor mass is concentrated near the minimal energy configuration $ \vec{\bar{q}} $, TT-cross may miss the associated sharp peak in $\ten{F}$, leading to large errors; 
(ii) TT-cross uses algorithms for sampling of tensor fibers; they may not resolve the peak unless the grid aligns well with $\vec{\bar{q}} $. These problems, related to sharply concentrated probability densities, were highlighted in recent works on TT inverse Rosenblatt transport maps \cite{cui2023scalable}. Consequently, applying standard TT-cross to the Boltzmann factor frequently results in tensor train decompositions with undesirably high ranks or large errors.
To resolve these issues, we sample the integrant based on Boltzmann factor's symmetry.

\textit{Discretization of the Boltzmann factor}---Assuming that the exponential function, $f(\vec{q})$, favors a unique dominant peak at $\qbar = [\bar{q}_1,\ldots,\bar{q}_{3N}]$,  the integrand
will be maximized at these coordinates.
To calculate \( Z_N \), we apply a cubature rule. For each dimension \( k=1,\ldots,3N \), we choose the integration domain \( [a_k, b_k] \) to be symmetric around \( \bar{q}_k \).
This choice generates a grid of size \( n^{3N} \), with \( \vec{\bar{q}} \), at its center. Importantly, due to the translational 
 and permutational invariance of the potential, the resulting discretization 
 of $\ten{F}$ on this grid displays a super-diagonal dominance. 
Specifically, for all indices $i$, $\ten{F}(i,i,...,i)= f(\bar{q}_1,\bar{q}_2,...,\bar{q}_{3N})=\bar{f}$, where the coordinates along the super-diagonal correspond to configurations preserving the relative positions of particles in the system. 

\begin{figure}[httb]
    \centering
    \includegraphics[width=0.6\linewidth]{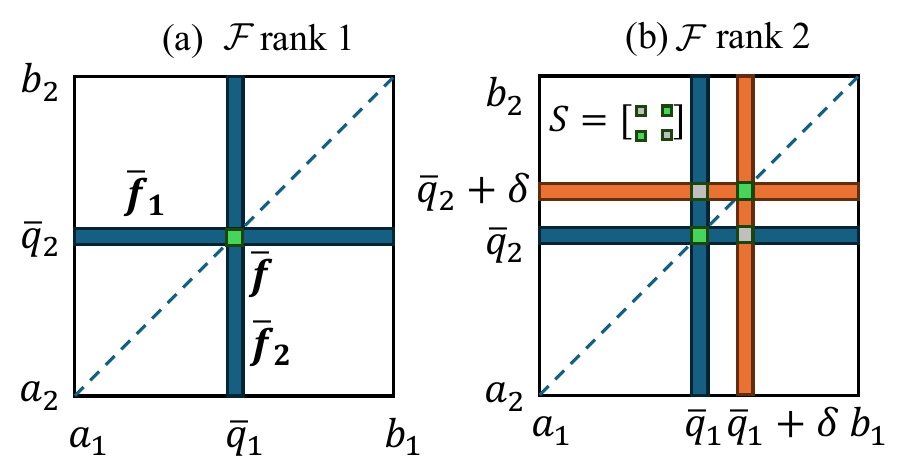}
    \caption{TT-cross interpolation of the Boltzmann factor. Panel (a) Rank-1 and Panel (b) Rank-2 .}
    \label{fig:cI_12}
\end{figure}

\textit{Rank-1 TT-cross of the $Z_N$ integrand}---In Figure~\ref{fig:cI_12} panel (a), we illustrate a rank-1 TT-cross for a 1D system, with two particles with coordinates \(\vec{q} =[q_1, q_2] \).
 We assume that the system possesses a global minimum at \(\bar{\vec{q}} =[\bar{q}_1, \bar{q}_2] \), maximizing the integrand to \(\bar{f} = f(\bar{\vec{q}})\). Integration is performed over $[a_1,b_1]\times[a_2,b_2]$, chosen to be symmetric around $[\bar{q}_1, \bar{q}_2]$. For rank-1, the maximum volume is reduced to the task of finding the maximum value of the function $f(\vec{q})$.
This value, $\bar{f}$ occurs at the cross section of two specific fibers of the tensor $\ten{F}$. The blue columns in Figure~\ref{fig:cI_12} panel (a), represent these two fibers: $\bar{\vec{f}}_1$ and $\bar{\vec{f}}_2$ of the 2D tensor $\ten{F}$. We formulate the following result (see the proof in End Matter,  \textit{Appendix B}):

\emph{Let the integrand \( f(\vec{q}) = e^{-\beta U(\vec{q})} \) attain its global maximum at \( \bar{\vec{q}} \), which corresponds to the multi-index \( \vec{j_1} = [j_1^1, \ldots, j_1^{3N}] \) on the super-diagonal of the tensor $\ten{F}$, and $j_1^1 = \ldots = j_1^{3N}$. Let \( \bar{f} := \mathcal{F}(\vec{j}_1) \) denote its value at that point.}
\emph{For each dimension \( k\), we define the fiber \( \bar{\mathbf{f}}_k \in \mathbb{R}^{n} \) as the \( k\)-mode fiber passing through the index \( \vec{j}_1 \). This set of fibers, \( \bar{\mathbf{f}}_k(i_k) = \mathcal{F}(j_1^{1}, \ldots, j_1^{k-1}, i_k, j_1^{k+1}, \ldots, j_1^{3N}) \) for $i_k = 1,\ldots,n$, intersects at $\vec{\bar{q}}$ with indices $\vec{j_1}$.
Then, the rank-1 TT-cross interpolation of $\mathcal{F}$ is:}
\begin{equation}
\hat{\mathcal{F}}= \mathbf{\bar{f}_1}\dfrac{1}{\bar{f}}\circ\mathbf{\bar{f}_2}\dfrac{1}{\bar{f}}\circ \ldots \circ \mathbf{\bar{f}_{3N}} = (\bar{f}^{-3N+1})\circ_{k=1}^{3N} \mathbf{\bar{f}_k},
\label{eqn:TT rank-1}
\end{equation}
\emph{where $\circ$ denotes tensor product, and $Z_N$ is,}
\begin{equation}
    Z_N \approx (\bar{f}^{-3N+1}) \prod_{k = 1}^{3N}(\Vec{w}_k^T\mathbf{\bar{f}_k}).
\label{eqn:Znrank-1}
\end{equation}

For crystal structures with identical particles, symmetry reduces computational complexity. In perfectly ordered solids, lattice periodicity makes all particles symmetrically equivalent, allowing us to calculate just one representative fiber, which yields:
\begin{equation}
    Z_N \approx (\bar{f}^{-3N+1})\prod_{k=1}^3(\Vec{w}_k^T\vec{\bar{f}_k)}^{N}.
\end{equation}

For crystal structures, such as Face-Centered Cubic (FCC), Body-Centered Cubic (BCC) or Diamond Cubic (DC), where the $x$, $y$, and $z$ directions are symmetrically equivalent,  only a single integral is needed:
\begin{equation}
    Z_N \approx (\bar{f}^{-3N+1})(\Vec{w}_1^T\vec{\bar{f}_1)}^{3N}.
\end{equation}

\textit{Rank-2 TT-cross of the $Z_N$ integrand}---In Figure~\ref{fig:cI_12} panel (b), we illustrate our approach. Again we choose  the integration domain to be symmetric around the minimum of the potential $[\bar{q}_1, \bar{q}_2]$. To calculate the integral we will compute TT-cross of $f(\vec{q})$ in this domain.
We select two points on the super-diagonal at distance $\delta$, where each point has the same value, $\bar{f}$. 
These two points, ${\bar{\bf q}} = \{\bar{q}_1,\bar{q}_2\}$, and the corresponding two points $\vec{q}^\delta_{2} = \{\bar{q}_1+\delta,\bar{q}_2\}$ outside the super-diagonal determine four fibers whose intersection forms the matrix $\mat{S}$. 
\begin{figure}[t]
    \centering
    \includegraphics[width=1.0\linewidth]{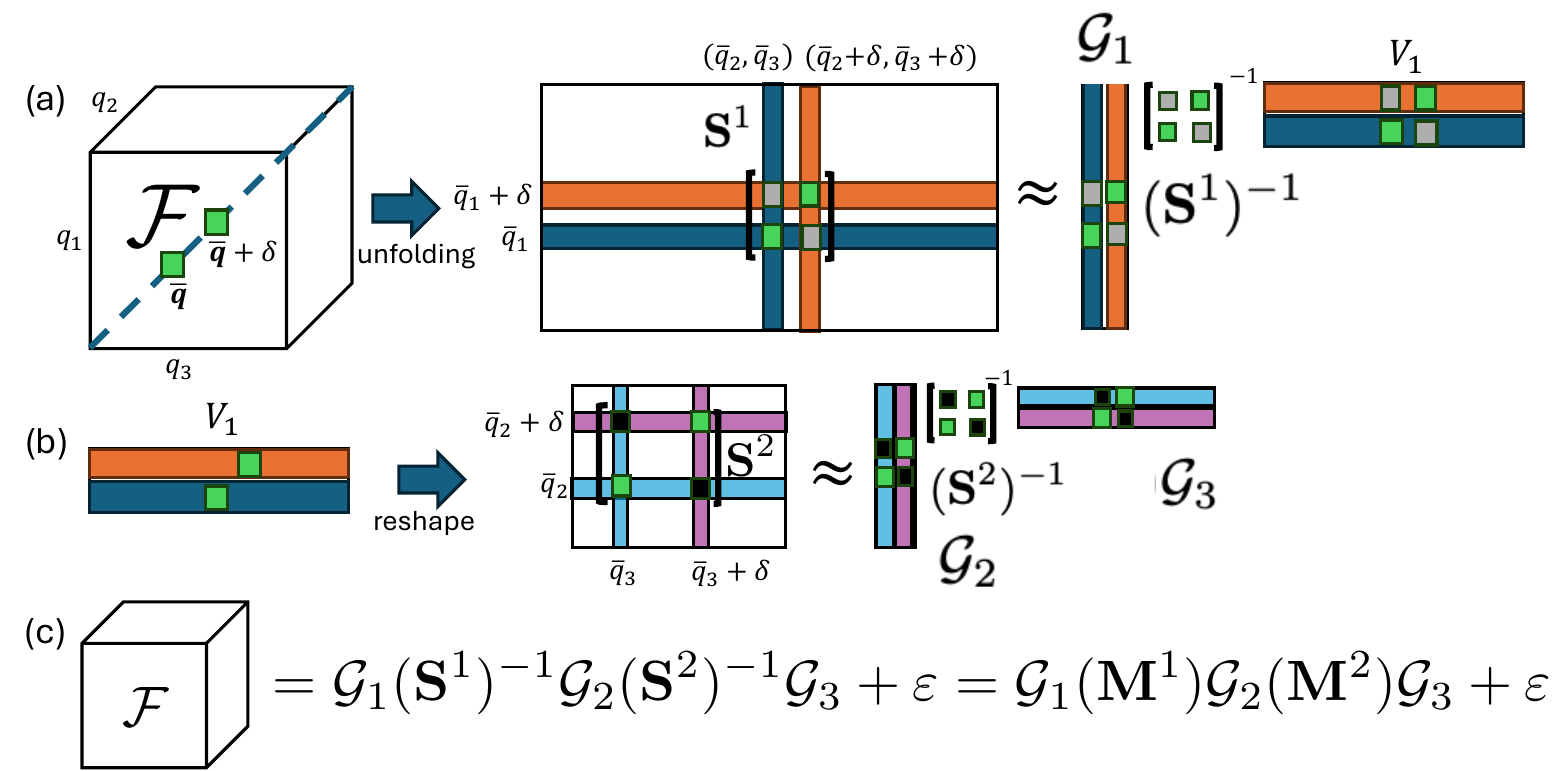}
    \caption{Illustration of a rank-2 TT format of  \( \mathcal{F} \) for a three particle system in 1D. 
    (a) The tensor is unfolded along its first dimension, and the rank-2 approximation is defined by selecting two points (green) on the super-diagonal. These two points determine four fibers (orange and blue), which intersect at four points defining: \( \ten{G}_1 \), \( (\mat{S^1})^{-1} \), and \( \mat{V}_1 \). 
    (b) \( \mat{V}_1 \) is reshaped, and the next \textbf{CUR}'s components are extracted. 
    (c) The final TT format is: \( \mathcal{F} = \ten{G}_1 \mat{M}^1 \ten{G}_2 \mat{M}^2 \ten{G}_3  + \varepsilon \) where \( \mat{M}^1 := (\mat{S^1})^{-1} \) and \( \mat{M}^2 := (\mat{S^2})^{-1}\).}
    \label{fig:rank2-decomp}
    \vspace{-0.7cm}
\end{figure}

Assuming that $f(\vec{q})$ is a very steep function,
than out of the diagonal: 
$f(\mathbf{q}^\delta_{1}) \approx f(\mathbf{q}^\delta_{2}) \approx 0$, and we have $|\det (\mat{S})| \approx \bar{f}^2$. Hence, the choice of this submatrix is close to the theoretical maximum volume, we need to calculate for the rank-2 TT-cross, and we can formulate the following result:

\emph{For N identical particles, let the point $\vec{\bar{q}}$ and point $\vec{\bar{q}} + \delta$ are on the super-diagonal. 
 Let $\mat{S}^k$, for $k = 1,\ldots,3N-1$, denote the submatrix of the k$^{th}$ subsequent unfolding of $\ten{F}$. Then, for a very steep function $f(\vec{q})$, $|\det (\mat{S}^k)| \approx \bar{f}^2$, which in many cases is close to the maximum volume we need to calculate for the rank-2 TT-cross interpolation.} 
 
 This rank-2 approach for TT-cross of the Boltzmann factor is illustrated in details for three particles in 1D, in Figure \ref{fig:rank2-decomp}, and the method for calculation of the TT-cores is given in End Matter, \textit{Appendix C}.

\begin{figure}
    \centering
    \includegraphics[width=0.6\linewidth]{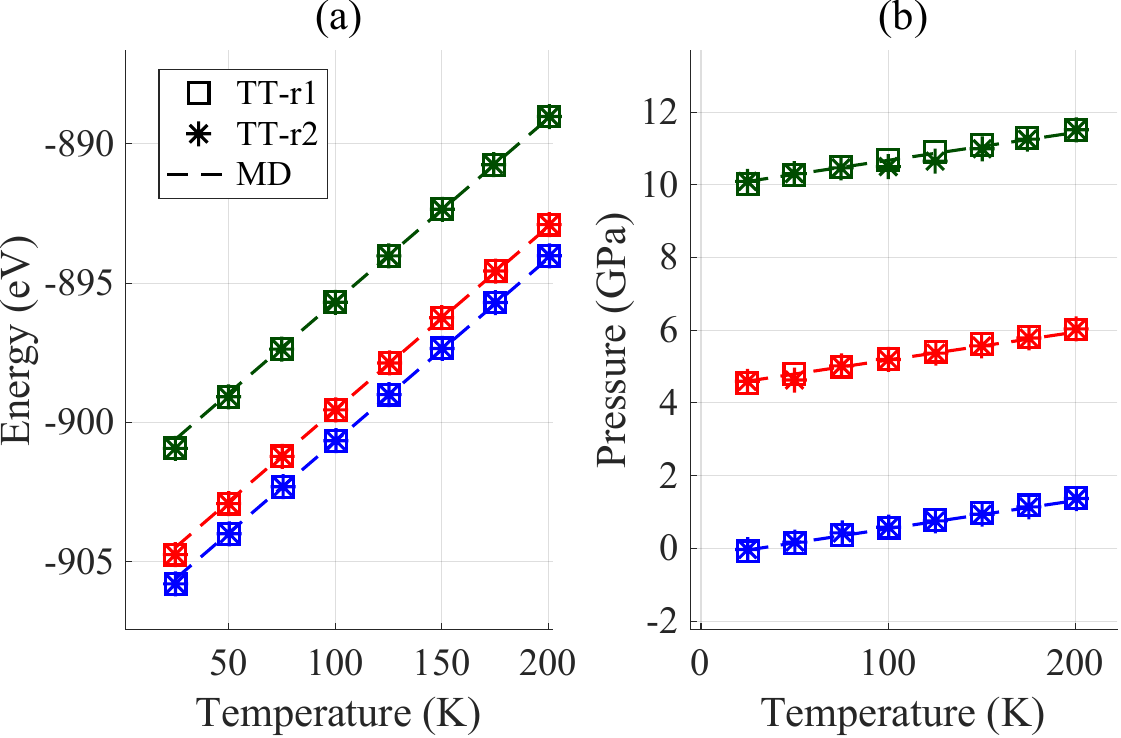}
    \caption{TT-rank 1 (squares) and TT-rank 2 (stars) predictions compared with MD results for three densities $9.49$ (green), $9.21$ (red), and $8.94$ (blue) $g/cm^3$, for $Cu$, calculated with tight-binding SMATB potential. Both TT-rank 1 and TT-rank 2 approximations of the internal energy (a) and pressure (b) are in excellent agreement with MD results.
    }
    \label{fig:Cu}
\end{figure}

\begin{figure}
    \centering
    \includegraphics[width=0.55\linewidth]{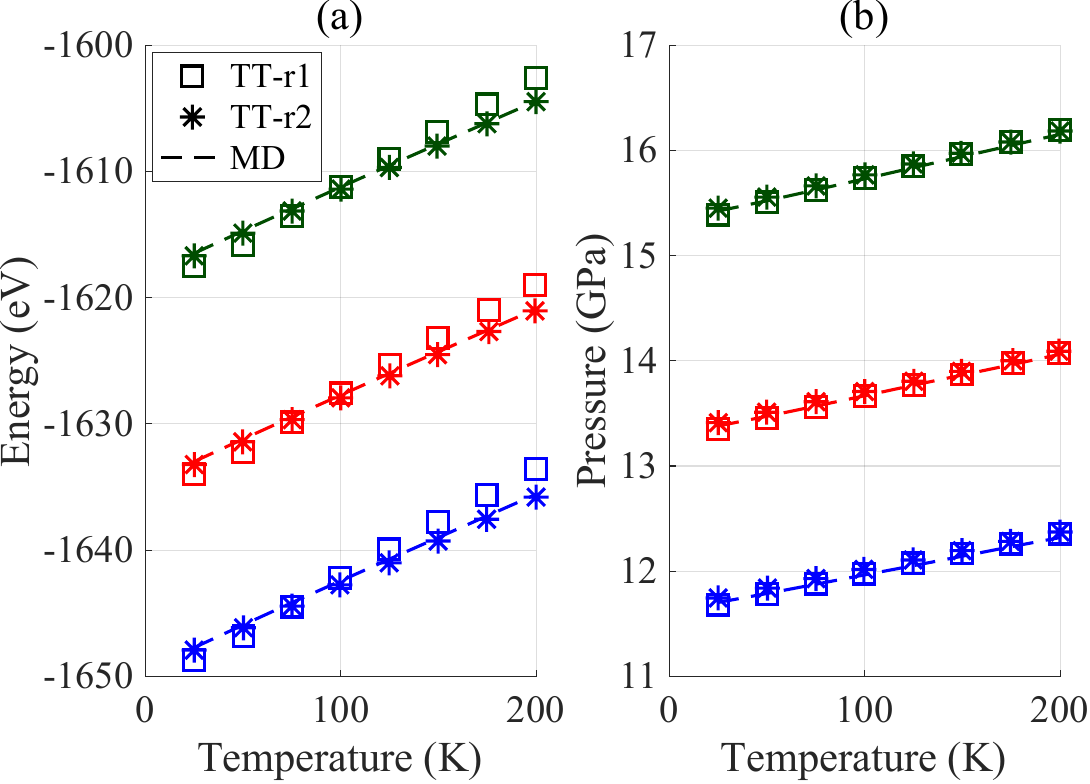}
    \caption{TT-rank 1 (squares) and TT-rank 2 (stars) predictions compared with MD results for FCC Argon calculated with the HIP-NN potential across three densities $3.48$ (green), $3.35$ (red), and $3.23$ (blue) $g/cm^3$. Rank-2 TT approximation of the internal energy (a) and pressure (b) consistently improves the approximation, yielding closer agreement with MD results. 
    }
    \label{fig:Ar-reg-HIPNN}
\end{figure}

\textit{Numerical Experiments}---It is well known from statistical mechanics that for a system of $N$ particles in a constant volume $V$, the mean internal energy of a system in the canonical ensemble, at temperature $T$, is given by: $E=-\frac{\partial \ln Q_N}{\partial \beta} = \frac{3N}{2\beta} - \frac{\partial \ln Z_N}{\partial \beta}$. Numerically, we can approximate $E$, by: $E \approx \frac{3N}{2\beta} - \frac{ \ln Z_N (\beta+\Delta\beta) - \ln Z_N(\beta)}{\Delta\beta}$. The thermodynamic pressure, of the considered phase, at volume $V$ and temperature $T$, is approximated by: $P=\frac{\partial \ln Q_N}{\beta\partial V} = \frac{\partial \ln Z_N}{\beta\partial V}\approx \frac{\ln Z_N(V+\Delta V) - \ln Z_N(V)}{\beta\Delta V}$. Both \( \Delta V \) and \( \Delta \beta \) are chosen empirically by gradually reducing their values until the computed internal energy and pressure become numerically stable.
To validate and demonstrate our approach, we calculate in two different ways these two thermodynamical quantities: (a) we calculate them by Molecular Dynamics (MD) simulations, and (b) by our TT-cross method. The MD parameters, are given in End Matter \textit{Appendix D}. We consider two materials systems: (a) Copper ($Cu$) and (b) Argon ($Ar$). For $Cu$, see Fig. (\ref{fig:Cu}), we use tight-binding potential \cite{cleri1993tight} that requires a rank-1 decomposition. For $Ar$, Fig. (\ref{fig:Ar-reg-HIPNN}), we use a Hierarchically Interacting Particle Neural Network (HIP-NN) many-body machine learning (ML) potential \cite{lubbers2018hierarchical}, trained with an MP2 dataset generated in End Matter \textit{Appendix E}. Interestingly, accurately reproducing the non-pairwise HIP-NN MD simulations required a rank-2 decomposition, in contrast to the other potentials examined in this paper which required only a rank-1 decomposition. For each structure with a given density, we fix the lattice constant and determine the coordinates of each particle using Atomic Simulation Environment (ASE) ~\cite{larsen2017atomic}. Next, for each structure with periodic boundaries conditions, we calculate the potential, using the Large-scale Atomic/Molecular Massively Parallel Simulator (LAMMPS) \cite{LAMMPS}, and evaluate the elements of the chosen fibers of $\ten{F}$.

\begin{figure}[]
    \centering
    \includegraphics[width=0.6\linewidth]{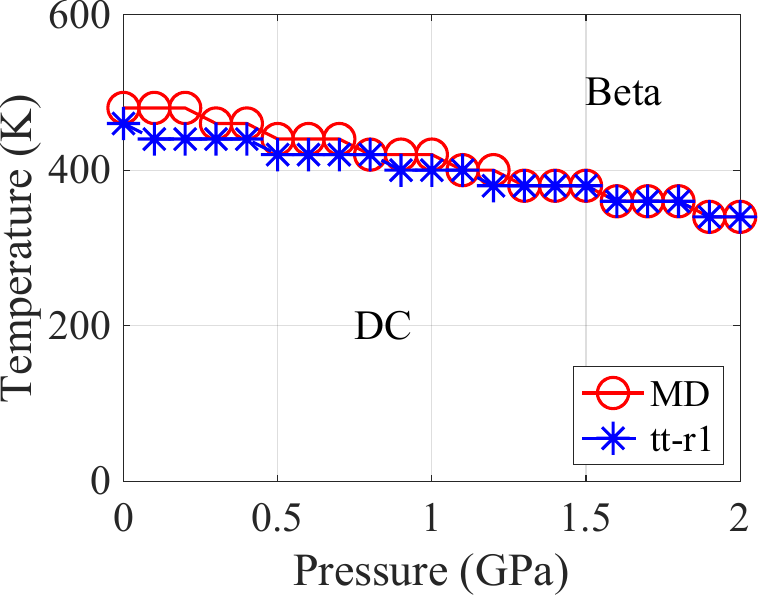}
    \caption{Phase diagram of Tin (Sn) using a MEAM potential, showing the \( \alpha \)-\( \beta \)-Sn transition. 
    The rank-1 TT (blue stars) reconstructs the phase boundary in close agreement with MD results (red circles).
    }
    \label{fig:Sn_MEAM_phase}
    \vspace{-0.7cm}
\end{figure}

Finally, we compare the computed phase transition temperature between $\alpha-Sn$ (Diamond Cubic) and $\beta-Sn $ (Body Centered Tetragonal) between a MD based thermodynamic integration and the TT-cross methodology. For this test a Modified Embedded Atom Method (MEAM) potential~\cite{lee2001second} was utilized due to the large number of simulations required to generate the themodynamic integration phase transition line.  Specifically, the MEAM potential developed by Ko and colleagues\cite{SnPotential} is known to accurately model the $\alpha-Sn$ to $\beta-Sn$ phase transition temperature. Fig. (\ref{fig:Sn_MEAM_phase}) shows that our TT-cross method reproduces the solid-solid phase transition between the diamond-cubic,  $\alpha - \mathrm{Sn(DC)}$ and \( \beta - \mathrm{Sn(\beta)} \) crystal structures. 
To create the initial configurations of $\mathrm{Sn(\beta)}$ or $\mathrm{Sn(DC)}$, at a given pressure, we run MD for fixed: crystal structure, number of particles, and volume. The MD simulations for the phase diagram used Calphy ~\cite{menon2021automated}, more details  are given in End Matter \textit{Appendix D}. As part of our method, for each phase and temperature,
we compute the configurational integral \( Z_N \) and thermodynamic pressure $P$, and evaluate the Gibbs free energy, \( G = (-\frac{1}{\beta} \ln Z_N + PV)/N \). The phase boundary is located at the temperature at which \( G_{\text{DC}} \approx G_{\beta} \). Our method required about 5 core hours whereas the MD simulations required 2560 core hours to complete.

\textit{Conclusions}---The statistical mechanics of strongly interacting atomic and molecular systems exemplifies how innovative tensor networks techniques can yield substantial benefits. Central to these systems is the configurational integral, a quantity that encapsulates the essence of particle interactions, which 
historically been deemed analytically and numerically intractable due to the curse of dimensionality.

In this paper, we demonstrate that our TT approach successfully evaluate configurational integrals in crystalline solids from first principles, enabling an accurate, fast, and efficient computation of thermodynamic functions, such as Helmholtz free energies and equations of state. 
In addition, our approach supports the use of more realistic ML potentials, thereby capturing the underlying physics more accurately than the pairwise additivity approximation. 

While the presented results focus on crystalline solids with identical particles, we are optimistic that further analysis of Boltzmann integrands and its symmetries will enable extension of our method. 

\textit{Acknowledgments}---We acknowledge
the support of NNSA for the U.S. DOE at LANL under Con-
tract No. DE-AC52-06NA25396, and Laboratory Directed
Research and Development (LDRD) for support through
20230067DR. LANL is managed by Triad National Security,
LLC, for the National Nuclear Security Administration of the
U.S. DOE under Contract 89233218CNA000001.

\bibliographystyle{apsrev4-2}
\bibliography{abbr}
\clearpage
\appendix

\newpage
\onecolumngrid
\begin{center}
\textbf{\large End Matter}
\end{center}
\twocolumngrid


\textit{Appendix A: The $\mat{CUR}$ approximation:}
The $\mat{CUR}$ is a matrix decomposition which approximate the matrix $\mat{A}$ by the product of subset of $r$ of its columns $\mat{C}$ and rows $\mat{R}$, and the pseudo-inverse of their intersection $\mat{U}$: $\mat{A} \approx \mat{C U R}$. In block-matrix form,
\begin{equation}
\label{eq:CUR}
\begin{bmatrix}
\mat{A}_{11}^{r \times r} & \mat{A}_{12} \\
\mat{A}_{21} & \mat{A}_{22}
\end{bmatrix}
\approx
\begin{bmatrix}
\mat{A}_{11} \\
\mat{A}_{21}
\end{bmatrix}
\mat{A}_{11}^{-1}
\begin{bmatrix}
\mat{A}_{11} & \mat{A}_{12}
\end{bmatrix} = \mat{CUR} \nonumber
\end{equation}
 
The $r$ columns and rows can be selected by the principle of maximum volume (maxvol) \cite{goreinov2001maximal}, that is, through searching for a submatrix $\mat{A}_{11}$ with maximum absolute determinant. 

The error of the approximation can be evaluated by comparison with SVD. The SVD which is finding the vectors (typically obtained after truncating the SVD to some small number $r$ of terms) providing the best rank-$r$ approximation to the matrix $\mat{A}$. If $ r \leq R = \text{rank}(\mat{A})$ and we define $\mat{A}_r = \sum_{i=1}^{r} \sigma_iu_iv_i^T$, where $\sigma_i, u,$ and $v$ are the singular values and singular vectors, then ||$\mat{A}$ – $\mat{A}_r$||$^2_F$ = min||$\mat{A}$ – $\mat{X}$||$^2_F$, for all $\mat{X}$ with \text{rank}$(\mat{X})\leq r$. Drineas, Kannan, and Mahoney constructed an additive-error $\mat{CUR}$ matrix decomposition by randomly construct a matrix $\mat{C}$ of columns, a matrix $\mat{R}$ of rows, and a matrix $\mat{U}$ such as, ||$\mat{A}$ – $\mat{CUR}$||$^2_F$ $\leq$ ||$\mat{A}$ – $\mat{A}_r$||$^2_F$ + $\varepsilon$||$\mat{A}$||$^2_F$, where $\varepsilon$ is a small number. The error of $\mat{CUR}$ approximation depends on the rank $r$, which is related to the rank $r$ of the truncated SVD of $\mat{A}$. Thus, when the spectrum of $\mat{A}$ decays slowly, i.e., the singular values \( \sigma_k \) for \( k > r \) are not neglectable, the accuracy requires a higher rank $r$, and rank-adaptive algorithms \footnote{\url{https://github.com/oseledets/TT-Toolbox}}, such as DMRG \cite{oseledets2011dmrg} and Amen-cross \cite{dolgov2014alternating}, are used in TT-cross.

\textit{Appendix B: Proof of the result for rank-1 TT-cross;}
To show this result we begin by unfolding the tensor \( \mathcal{F} \) along the first mode to obtain the matrix \( \vec{F}_1 \in \mathbb{R}^{n \times n^{3N-1}} \), where each entry is indexed by \([i_1, I_{i_2 i_3 \ldots i_{3N}}]\), and \( I_{i_2 i_3 \ldots i_{3N}} \) is a flattened multi-index combining the remaining \( 3N-1 \) dimensions.

According to the maximum volume principle, the best rank-1 approximation of \( \vec{F}_1 \) that passes through the pivot index \( [j_1^1, I_{j_1^2 j_1^3 \ldots j_1^{3N}}] \) is given by:
\[
\vec{F}_1 \approx \vec{F}_1(:, I_{j_1^2 j_1^3 \ldots j_1^{3N}}) \cdot \frac{1}{\bar{f}} \cdot \vec{F}_1(j_1^1, :) = \bar{\mathbf{f}}_1 \cdot \frac{1}{\bar{f}} \cdot \vec{F}_1(j_1^1, :).
\]

We then reshape the row vector \( \vec{F}_1(j_1^1, :) \in \mathbb{R}^{1 \times n^{3N-1}} \) into a matrix \( \vec{F}_2 \in \mathbb{R}^{n \times n^{3N-2}} \), corresponding to unfolding along the second mode. Applying the maximum volume rank-1 approximation again at index \( j_1^2 \), we obtain:
\[
\vec{F}_2 \approx \vec{F}_2(:, I_{j_1^3 j_1^4 \ldots j_1^{3N}}) \cdot \frac{1}{\bar{f}} \cdot \vec{F}_2(j_1^2, :) = \bar{\mathbf{f}}_2 \cdot \frac{1}{\bar{f}} \cdot \vec{F}_2(j_1^2, :).
\]

We continue this sequential procedure, reshaping and applying the max-volume rank-1 decomposition through the fixed multi-index \( \vec{j}_1 \) at each unfolding. After repeating this process up to the \( (3N-1)^\text{th} \) mode, we arrive at the final rank-1 TT-cross interpolation of \( \mathcal{F} \) as described in Equation~\eqref{eqn:TT rank-1}:
\[
\hat{\mathcal{F}} = \bar{f}^{-3N+1} \circ_{k=1}^{3N} \bar{\mathbf{f}}_k.
\]

Finally, applying the TT cubature rule (Eqn.~\eqref{eq:TT - cubature}) to the rank-1 TT format yields the approximate integral \( Z_N \) in Eqn.~\eqref{eqn:Znrank-1}:
\[
Z_N \approx \bar{f}^{-3N+1} \prod_{k=1}^{3N} \left( \vec{w}_k^T\bar{\mathbf{f}}_k \right).
\]
\textit{Appendix C: Calculating the TT cores in rank-2 approximation:} Assuming that the indices of the point $\vec{\bar{q}}$ are $\vec{j_1} := [j_1^1,j_1^2,\ldots,j_1^{3N}]$, and of $\vec{\bar{q}} + \delta$ are $\vec{j_2} := [j_2^1,j_2^2,\ldots,j_2^{3N}]$, the cores of the rank-2 TT decomposition can be calculated as:\\
The elements of $2\times2$ matrices $\vec{S}^k$ are calculated as:
\begin{align*}
    \vec{S}^k_{1,1} &=  \ten{F}(j_1^1,\ldots,j_1^k,j_1^{k+1},\ldots,j_1^{3N}),\\
    \vec{S}^k_{1,2} &=  \ten{F}(j_1^1,\ldots,j_1^k,j_2^{k+1},\ldots,j_2^{3N}),\\
    \vec{S}^k_{2,1} &=  \ten{F}(j_2^1,\ldots,j_2^k,j_1^{k+1},\ldots,j_1^{3N}),\\
    \vec{S}^k_{2,2} &=  \ten{F}(j_2^1,\ldots,j_2^k,j_2^{k+1},\ldots,j_2^{3N}).
\end{align*}
with $\vec{M}^k = [\vec{S}^k]^{-1}$.\\
The elements of the fibers of the 3D TT-cores $\ten{G}_1(i,:,j)$ and  $\ten{G}_{3N}(i,:,j)$ are calculated as,
\begin{align*}
    \ten{G}_1(1,:,1)&= \{\ten{F}(i_1,j_1^2,...,j_1^{3N})\}^n_{i_1=1},\\
    \ten{G}_1(1,:,2)&= \{\ten{F}(i_1,j_2^2,...,j_2^{3N})\}^n_{i_1=1},\\
    \ten{G}_{3N}(1,:,1) &= \{\ten{F}(j_1^1,...,j_1^{3N-1},i_{3N})\}^n_{i_{3N}=1},\\
    \ten{G}_{3N}(2,:,1)&= \{\ten{F}(j_2^1,...,j_2^{3N-1},i_{3N})\}^n_{i_{3N}=1},
\end{align*} 
and  for  $\ten{G}_k(i,:,j),\ k \in \{2, . . . ,3N-1\}$,
\begin{align*}
\ten{G}_k(1,:,1)&= \{\ten{F}(j_1^1,\ldots,j_1^{k-1},i_k,j_1^{k+1},\ldots,j_1^{3N})\}_{i_k =1}^n,\nonumber\\
\ten{G}_k(2,:,1)&= \{\ten{F}(j_2^1,\ldots,j_2^{k-1},i_k,j_1^{k+1},\ldots,j_1^{3N})\}_{i_k =1}^n,\nonumber\\
\ten{G}_k(1,:,2)&= \{\ten{F}(j_1^1,\ldots,j_1^{k-1},i_k,j_2^{k+1}, \ldots,j_2^{3N})\}_{i_k=1}^n,\nonumber\\
\ten{G}_k(2,:,1)&= \{\ten{F}(j_2^1,\ldots,j_2^{k-1},i_k,j_1^{k+1},\ldots,j_1^{3N})\}_{i_k =1}^n.\nonumber
\end{align*}

If the system is formed by one type of particles with periodically boundary conditions, and the $x$, $y$, and $z$ directions are symmetrically equivalent, such as, in Face-Centered Cubic (FCC), Body-Centered Cubic (BCC) or Diamond Cubic (DC) structures, all the matrix $\vec{M}^k$, and all TT-cores $\ten{G}_k; k \in \{2, . . . ,3N-1\}$ are identical.
The above calculations can be easily generalized to a rank-$r$ TT-cross, by choosing a set of $r$ points on the super-diagonal and constructing the fibers of the 3D TT-cores, $\ten{G}_k$, and the elements of the $r \times r$ matrices $\mat{M^k}$ as described above.\\
\textit{Appendix D: Molecular Dynamic Procedures:}
\begin{itemize}
\item \textit{Argon and Copper Molecular dynamics (MD) simulations:} The MD are performed using LAMMPS to study FCC Argon and Copper at various densities under fixed temperature conditions. A \( 4 \times 4 \times 4 \) FCC supercell is generated using the ASE package ~\cite{larsen2017atomic} and rescaled to match the target density. The SMATB (Second Moment Approximation to Tight Binding) potential~\cite{rosato1989thermodynamical} is used for Copper, while Argon is modeled using the Hierarchically Interacting Particle Neural Network (HIP-NN) machine learning potential ~\cite{hipnnTen}. A small custom training dataset for the Argon HIP-NN potential was generated through active learning, as described in Appendix E. Simulations are conducted in the NVT ensemble using a Nosé--Hoover thermostat with a timestep 0.2~fs, and a total run length of 20,000 steps.
\item \textit{Tin Phase Diagram:}
The phase diagram of tin $Sn$ is computed using the Calphy package, an open-source Python framework for automated free energy and phase boundary calculations~\cite{menon2021automated}. Calphy employs the thermodynamic integration method combined with molecular dynamics simulations to evaluate the Gibbs free energy of different crystalline phases. In this study, the Diamond Cubic (DC), $\alpha$ and \( \beta \)-Sn phases are considered. For these simulations, Calphy was used in "temperature sweep" mode to scan Gibbs free energy over a variety of different initial pressures. Specifically, a simulation was run every 0.05 GPa from 0 to 2 GPa from 100K to 600K starting in both the $\alpha-Sn$ and $\beta-Sn$ configuration. 10000 equilibration steps were performed before running the temperature sweep over 50000 steps, with a 1 fs MD time step. For each initial pressure, the temperature at which the $\alpha-Sn$ and $\beta-Sn$ Gibbs free energies crossed is reported as the phase transition temperature. These temperatures were then fit to a 25K grid, over which the TT-method was performed, since this method required a grid in both temperature and pressure space. 
\end{itemize}

\textit{Appendix E: An active learned Argon MP2 dataset:}
In order to train the Argon HIP-NN machine learning potential, a dataset of 425 MP2 calculations was generated via active learning. RI-MP2 implemented in Q-Chem\cite{Qchem17012015} utilizing the CC-PVTZ\cite{ccpvtz} basis set was utilized as the underlying reference method. Preliminary MD simulations of up to 40 Ar atoms were performed held together with a linear spherical well potential with a radius of 7.0 \AA and a force of 0.01 eV/\AA. These MD simulations were performed with an ensemble of 8 HIP-NN models, and stopped when the standard deviation of the force predictions exceeded 0.25 eV/\AA. After 20 active learning generations, a HIP-NN model with 60 atomic features, 2 interaction layers, 3 atomic layers, a soft cutoff of 4.0 \AA and a hard cutoff of 5.5 \AA was trained to the gathered data. A random split of 80\% training, 10\% validation, and 10\% testing was generated from the 425 structures. The resulting model had test set accuracies as reported in \ref{tab:HIPNNacc}

\begin{table}
    \centering
    \begin{tabular}{c|cc}
               & MAE & RMSE \\
        \hline
        Energy (eV) & 0.40 & 0.65 \\
        Force (eV/\AA) & 0.35 & 0.72 \\
    \end{tabular}
    \caption{Test set accuracies for the developed HIP-NN Argon model. Model was primarily used to demonstrate that higher order tensor integrations are required for modeling non-pairwise potential energy surfaces.}
    \label{tab:HIPNNacc}
\end{table}

\end{document}